# Parameters of the Dzyaloshinsky-Moriya type weak ferromagnetism for some perovskite compounds

**D. DIMITROV, I. RADULOV, V. LOVCHINOV, A. APOSTOLOV[a]**

*Institute of Solid State Physics, Bulgarian Academy of Sciences, 72 Tzarigradsko Chaussee Blvd., 1784 Sofia, Bulgaria.*
[a] *University of Sofia, Faculty of Physics, 5 J.Bourchier Blvd.,1164 Sofia, Bulgaria*

***Corresponding author's e-mail:*** dimitrov@issp.bas.bg

Magnetic properties of rare-earth antiferromagnetic mixed manganites, chromites and ferrites are investigated and the Dzyaloshinsky parameter is defined for them. The angle a, at which the magnetization curves separate from the antiferromagnetic axes is determined.

## 1. Introduction

Compounds with distorted perovskite structure of the 4-f and 3-d transition metals with the common formula $LnTO_3$ (where Ln is rare-earth element, T is an element from the Fe group) are the most multifold binary oxides of these two groups elements. Wide range of stability for this structure allows the realization of combinations of the Lanthanides with all the transition metals except the Nickel. Quite interesting physical phenomena take place in these oxides like charge and orbital ordering; relatively independent magnetic lattices of the both metals; particular magnetic structures; high optical indicators; giant magneto-resistance; peculiar dielectric and ferroelectric properties etc.

We have investigated the magnetic properties of pure ferrites and chromium-based materials as well as of some mixed type oxides like $Ho_xTb_{1-x}O_3$, $HoMn_xFe_{1-x}O_3$, $HoMn_xCr_{1-x}O_3$ and $DyFe_xCr_{1-x}O_3$.

The samples magnetic properties are investigated between 2 K and 1000 K. Magnetic fields up to 2.5 T and the ballistic method are applied below 300 K and a magnetic relaying balance is used above this temperature.

Aim of the investigation is to measure the magnetic properties in respectively wide temperature and magnetic regions in order to determine the Dzyaloshinsky-Moria effect which appears in the perovskites.

## 2. Results and discussion

The investigations show that above a certain transition temperature $T_{N1}$ all the compounds are in paramagnetic state and the measured susceptibility $\chi 1$ follows exactly the Curie-Weiss law with an effective magnetic moment near to the calculated of magnetic ions at the applied proportions and with negative asymptotic temperature $\theta 1$ (see fig.1). Negative value of this quantity indicates that the negative (anti-ferromagnetic) interactions predominate between the magnetic ions. Each of the measured magnetizations (see fig.2) between $T_{N1}$ and a certain lower transition temperature $T_{N2}$ is a sum of two terms: one which is proportionally changes to the applied magnetic field and characterizes a second susceptibility $\chi 2$; and second ferromagnetic term which allows measuring the spontaneous magnetization at each temperature. Temperature dependence of the reciprocal $\chi 2$ value is a straight line with declination different from that of $1/\chi 1$ and gives the opportunity to determine second negative paramagnetic temperature $\theta 2$ (see fig.1). At temperatures, e.g. lower than 15 K the susceptibility experimental points become

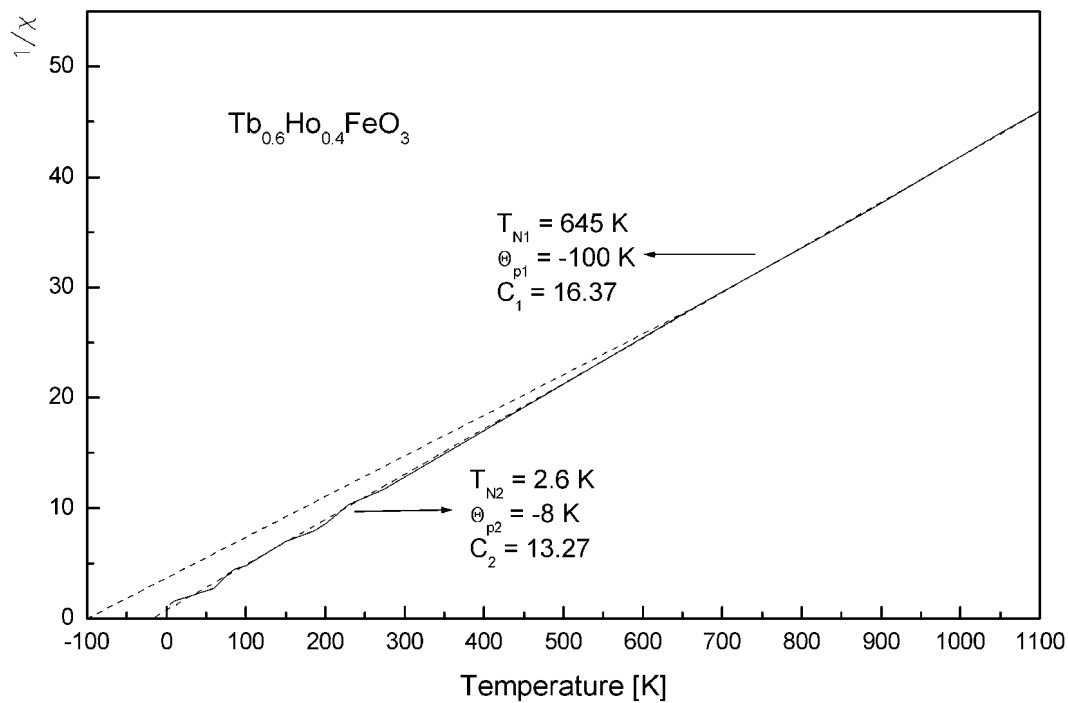

Figure 1. (Ho-Tb)FeO$_3$ susceptibility.

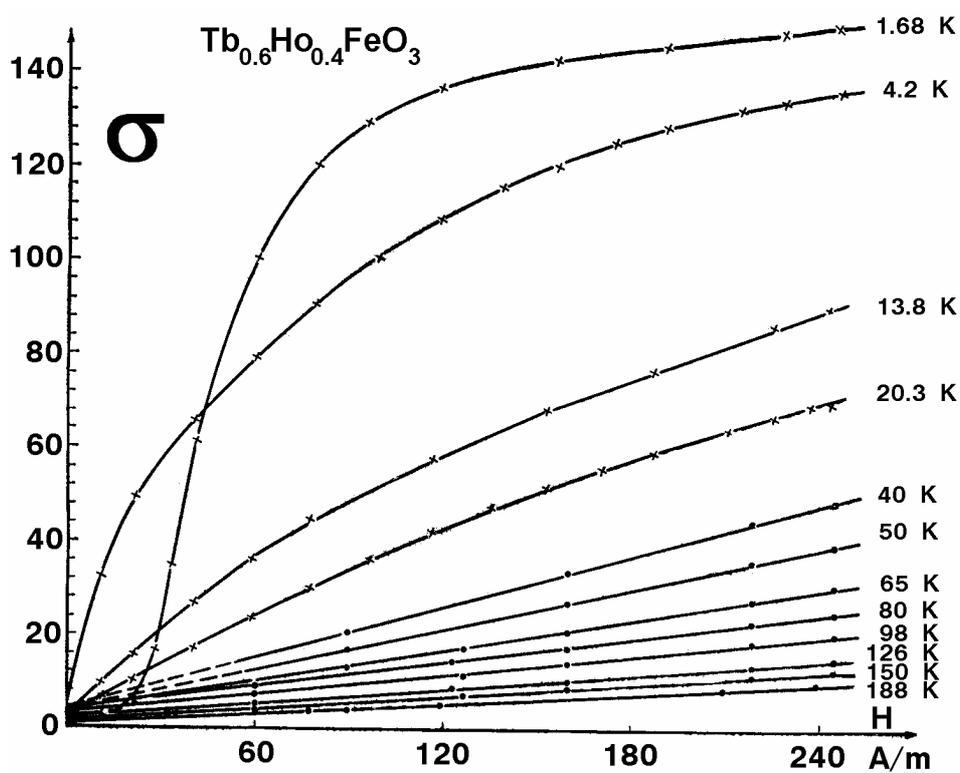

Figure 2. Magnetic dependences M = f(H) for (Ho-Tb)FeO$_3$ compound.

detached from the straight line $1/\chi_2$ and show phase transition which is taking place at different temperatures for the respective compounds (see table.1)

Total magnetization observed experimentally in the regarded area is:

$$M + M_{4f} + M_{3d} = M_{4f} + \chi_1.H + m_{3d} = \frac{c_{4f}}{T-\theta_{p2}}(H + n_{3d-4f}.m_{3d}) + \chi_{3d}.H + m_{3d},$$

where $\chi_{3d}$ presents the weak antiferromagnetic susceptibility of the arranged 3d metals, $c_{4f}$ is the Curie constant for the Lanthanide. It is much weaker than the paramagnetic susceptibility of the Lanthanides (this can be directly measured for compounds including Lu, La or Y) and, therefore, can be neglected. Thus, for the magnetization with reasonable approximation can be written:

$$M = \frac{c_{4f}}{T-\theta_{p2}}(H - n_{3d}.4f.\sin\alpha.M_{Fe}) + M_{3d}.\sin\alpha.$$

Spontaneous ferromagnetic magnetization measured at $H = 0$ fulfils the equation

$$\frac{M_{spon}}{M_{3d}}(T-\theta_{p2}) = n_{3d-4f}.\sin\alpha.c_{4f} + \sin\alpha\,(T - \theta_{p2}).$$

Here $m_{3d}$ is substituted by $M.\sin\alpha$ which is evident.

If in the left part of the equation $\theta_{p2}$ is replaced by $\theta_{pR.E.}$ ($\theta_{pR.E.}$ is the paramagnetic temperature for the rare-earth element) and it is plotted versus $(T - \theta_{pR.E.})$ the result should be a straight line. This is graphically verified. The declination of the line and the piece form the X-axis define the angle $\alpha$ and the coefficient $n_{4f-3d}$ of weak interaction between the Lanthanides and the rest transient elements. These results are presented in table 1. It is worth to notify that the angle $\alpha$ for the $HoFeO_3$ compound was directly identified by pulse neutronography [5] and the obtained value does not essentially differ from those given in the table 1.

| Compound | α (mrad) | n | D (erg/gauss) |
|---|---|---|---|
| $HoMn_{0.1}Fe_{0.9}O_3$ | 3 | -273 | 1.4 |
| $HoMn_{0.2}Cr_{0.8}O_3$ | 12 | -115 | 2.8 |
| $HoMn_{0.25}Cr_{0.75}O_3$ | 5 | -104 | 1.0 |
| $HoMn_{0.3}Cr_{0.7}O_3$ | 30 | -127 | 7.6 |
| $HoMn_{0.4}Cr_{0.6}O_3$ | 30 | -105 | 6.4 |
| $HoMn_{0.7}Fe_{0.3}O_3$ | 14 | -202 | 6.2 |
| $Ho_{0.6}Tb_{0.4}FeO_3$ | 18 | -326 | 4.0 |
| $HoFeO_3$ | 4 | -263 | 4.0 |
| $TbFeO_3$ | 6 | -368 | 4.4 |
| $DyFeO_3$ | 7 | -335 | 2.3 |
| $LuFeO_3$ | 17 | -210 | 7.1 |

We have applied an entire investigation to the lutetium and yttrium ferrite systems. Lutetium ferrite, for example, orders anti-ferromagnetically below 625 K with weak ferromagnetic moment associated with the anti-ferromagnetism. The ferromagnetic moment value is low (about 0.03 – 0.05 MB) and remains unchanged up to 20 K. Ratio between the experimentally determined weak ferromagnetic moment and the calculated by the Brilluen's law is constant and indicates that the angle of deflection of the Fe spontaneous magnetisation from the anti-ferromagnetic axis is constant at all the temperature region.

These results are checked out for the yttrium ferrite as well as using the available data for Lu and Y chromites in the literature [1, 2].

Explanation of the observed features [2, 3] is classic: At high temperatures the compounds are in paramagnetic state; under the first transition temperature only the 3d-lattice orders anti-ferromagnetically; the 4f-lattice (if the lanthanide is paramagnetic) retains in disordered state. The 3d-lattice anti-ferromagnetism is accompanied by weak ferromagnetic moment which is measured for La, Lu, and Y in pure mode. The weak ferromagnetic moment polarises the paramagnetic lanthanides and can induces a total ferromagnetic moment up to 3 MB (e.g. for Ho ferrite). The low angle $\alpha$ assessment (at which 3d magnetisations of different compounds decline from the anti-ferromagnetic axis) is one of the aims of the present work.

Here below we shall use the Neel model [4] concerning the molecular (self arranged) field applied to the complicated perovskite structure in the following way:
— 3d-lattice divides into two antiparallel lattices, each including half of the all atoms;
— Molecular field coefficients of the 3d-lattice are $n_{3d}$ and $n`_{3d}$. $n_{3d}$ accounts for the near neighbour interactions, i.e. between the both sublattices and $n`_{3d}$ accounts for the interactions between the magnetic atoms which are inside of each from the sublattices;
— Noncolinearity of the both antiferromagnetically situated sublattices leads to their diversion from the antiferromagnetic axis with a small angle $\alpha$ and therefore, leads to weak magnetic moment $m_{3d}$;
— Rare earth ions can be divided also into two antiparallel lattices with coefficients for them $n_{R.E.}$ and $n`_{R.E.}$, respectively.
— The molecular field coefficient $n_{3d-4f}$ defines the weak interaction between the magnetic lanthanides and 3-d ions.

Lanthanide ions are in paramagnetic state in the temperature region between $T_{N1}$ and $T_{N2}$ and are under the influence of the external magnetic field and the 3d-molecular field through the coefficient $n_{3d-4f}$:

$$M_{4f} = \frac{c_{4f}}{T - \theta_{p2}} (H + n_{3d-4f}.m_{3d}),$$

here $c_{4f}$ is the experimentally measured by the $1/\chi 2$ value of the Curie constant for the lanthanide (see table1).

The physical origin of the weak ferromagnetism, i.e. the noncolinearity of th 3d antiferromagnetic sublattices can result from:
— One ion anisotropy (classical example – NiFe) originates from the existence of two nonequivalent crystallographic nodes occupied by equivalent ions. Thus, each ion is affected by different crystallographic field and, therefore, nonparallel axes of anisotropy are created [6];
— Dzyaloshinsky asymmetric exchange (classical example - $\alpha$-$Fe_2O_3$) where the simultaneous action of the spin-orbital and mediate exchange junction leads to anisotropic interaction of the type D(M1 x M2), where D is vector power parameter and M1 and M2 are the magnetisations of the both antiferromagnetic lattices [7].

Gorodetsky et al. show for the first time [8] that the probable mechanism for the weak ferromagnetism of the rare-earth perovskites is of the second type and now it is commonly accepted opinion. It is of particular inerest to determine experimentally not only the angle $\alpha$ but the power parameter D (Dzyaloshinsky-Moriya parameter) also.

The energy level of the 3d-system in the examined compounds can be defined with reasonable approximation only from the isotopic and asymmetric exchange forces (the crystallography anisotropic energy is proportional to the $\sin\alpha$ and in our relatively crude model can be neglected). For the case it can be written:

$$W = -1/2 \sum_{i=1}^{2} h_i M_i - D[M1.M2]$$

$h_1 = n_{3d}M2 + n`_{3d}M1;$
$h_2 = n_{3d}M1 + n`_{3d}M2$

and thus $|M1| = |M2|$, $D \parallel (M1.M2)$, then

$W = n.M^2_{3d}.\cos2\alpha - DM^2_{3d}\sin2\alpha - 2n`_{3d}M^2;$

$dW/d\alpha = -n.\sin\alpha - D.\cos\alpha = 0$

$tg2\alpha = -D/n$

$|\alpha| = D/2n.$

There are few other direct measurements of this parameter which is of the order of ten miliradians as it is shown in table 1.

### 3. Conclusion

Parameters characterising the weak ferromagnetism associated to the antiferromagnetism in double oxides with perovskite structure of the 3d and 4f elements were determine applying direct magnetic measurements in wide temperature range.

### 4. Acknowledgements

This work is supported by the CGRI (*Commissariat Général aux Relations Internationales – Belgium*) and by the Bulgarian Academy of Sciences through a joint collaborative research program. The work of I. Radulov is supported by NATO EAP.RIG. 981824.